\documentclass{emulateapj}

\usepackage{psfig}
\usepackage{apjfonts}
\usepackage{times}

\def\pwn{G189.22+2.90}
\def\psr{J0617+2221}
\def\mousepsr{J1747--2958}
\def\mouse{G359.23--0.82}
\def\snr{IC~443}
\def\cxo{{\em Chandra}}
\def\xmm{{\em XMM-Newton}}

\def\kms{km~s$^{-1}$}
\def\etal{{\rm et~al.\ }}

\shorttitle{X-RAY STRUCTURE OF THE BOW SHOCK IN SNR~IC~443}
\shortauthors{GAENSLER ET AL.}

\begin{document}
\title{The X-ray Structure of the Pulsar Bow Shock G189.22+2.90
in the Supernova~Remnant~IC~443}
\submitted{To appear in {\em The Astrophysical Journal}}
\author{B. M. Gaensler,\altaffilmark{1,2,3} 
S. Chatterjee,\altaffilmark{4,1}
P. O. Slane,\altaffilmark{1}
E. van der Swaluw,\altaffilmark{5}
F. Camilo\altaffilmark{6} 
and J.~P.~Hughes\altaffilmark{7}}
\altaffiltext{1}{Harvard-Smithsonian
Center for Astrophysics, 60 Garden Street, Cambridge, MA 02138;
bgaensler@cfa.harvard.edu}
\altaffiltext{2}{Alfred P.\ Sloan Research Fellow}
\altaffiltext{3}{Present address: School of Physics A29, University
of Sydney, NSW 2006, Australia}
\altaffiltext{4}{Jansky Fellow, National Radio Astronomy Observatory, 520 Edgemont Road,
Charlottesville, VA 22903}
\altaffiltext{5}{Royal Netherlands Meteorological Institute (KNMI),
PO Box 201, 3730 AE De Bilt, The Netherlands}
\altaffiltext{6}{Columbia Astrophysics Laboratory, Columbia University,
  550 West 120th Street, New York, NY10027}
\altaffiltext{7}{Department of Physics and Astronomy, 
Rutgers University, 136 Frelinghuysen Road, Piscataway, NJ 08854}

\begin{abstract}

We present a deep observation with the {\em Chandra X-ray Observatory}\
of the neutron star bow shock \pwn\ in the supernova remnant
(SNR) \snr.  Our data confirm the cometary morphology and central point
source seen previously, but also reveal considerable new structure.
Specifically, we find that the X-ray nebula consists of two distinct
components: a ``tongue'' of bright emission close to the neutron star,
enveloped by a larger, fainter ``tail''. We interpret the tongue and
tail as delineating the termination shock and the post-shock flow,
respectively, as previously identified also in the pulsar bow shock
\mouse\ (``the Mouse''). However, for \pwn\ the tongue is much less
elongated than for the Mouse, while the tail is much broader. 
These differences are consistent with the low Mach number,
$\mathcal{M} \la 2$, expected for
a neutron star moving through the hot gas in a SNR's interior, supporting the
case for a physical association between \pwn\ and \snr.
We resolve the stand-off
distance between the star and the head of the bow shock, which allows
us to estimate a space velocity for the neutron star of $\sim230$~\kms,
independent of distance.  We detect thermal emission from the neutron
star surface at a temperature of $102\pm22$~eV, which is consistent
with the age of SNR~\snr\ for standard neutron
star cooling models.
We also identify two compact knots of hard emission
located $1''-2''$ north and south of the neutron star.

\end{abstract}


\section{Introduction}
\label{sec_intro}

The relativistic winds from energetic pulsars generate broadband
synchrotron emission when they interact with their surroundings.
New high resolution X-ray observations have revealed amazing details
in the structure of the resulting pulsar wind nebulae (PWNe)
\citep{wh06,gs06}. Most of these systems resemble the Crab Nebula,
and typify early phases of PWN evolution.
However, after a few thousand years a pulsar's motion becomes
supersonic with respect to its surroundings, and a bow shock is formed.
While only a small group of pulsar bow shocks have been as yet
identified, these sources provide an additional opportunity to
probe pulsar winds, using very different boundary conditions
from those experienced by ``Crab-like'' PWNe.

\begin{figure} 
%\plotone{f1.eps}
%\centerline{\psfig{file=fig_wide.eps,width=\textwidth,angle=270}}
\centerline{\psfig{file=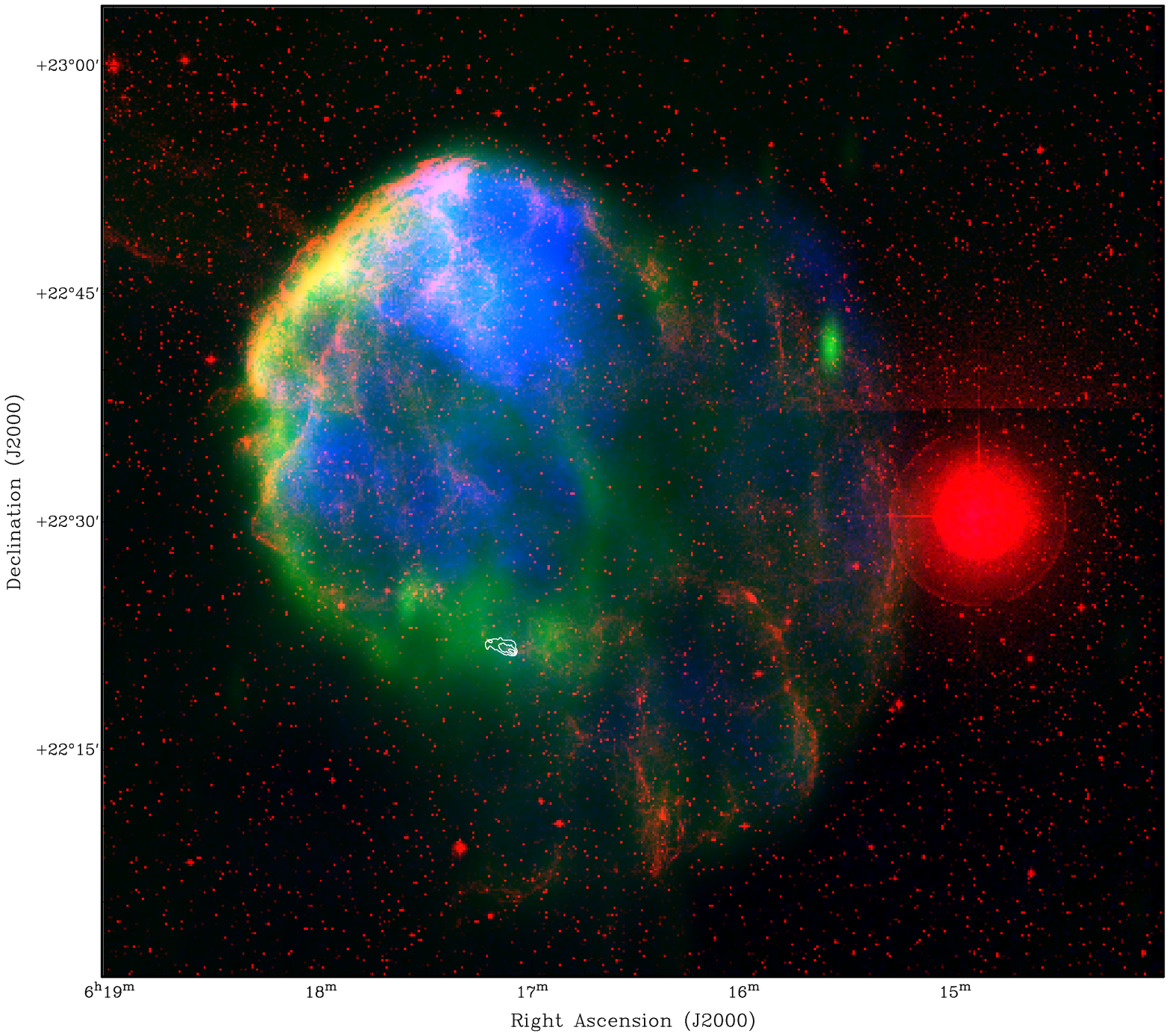,width=\textwidth,angle=270}} 
\caption{A multiwavelength view of the SNR~\snr\ and the PWN \pwn. Red
shows 670-nm emission from the Second Palomar Observatory Sky Survey;
green shows 1.4-GHz radio data taken by
the DRAO Synthesis Telescope \citep{lea04}; blue shows 0.1--2.4~keV X-ray
data taken by the {\em ROSAT}\ PSPC \citep{aa94}. The white contours
show 8.5-GHz Very Large Array data on \pwn\ at a resolution
of $9''\times 8''$ (O2001), with contour levels drawn at 20\%, 50\% and 80\% of
the peak of 3.7~mJy~beam$^{-1}$.  The bright star to the west of \snr\
is $\eta$~Gem.}
\label{fig_wide}
\end{figure}

\cite{gvc+04}, hereafter G2004, presented a detailed study of \mouse\
(``the Mouse''), a spectacular radio and X-ray bow shock powered by
the energetic pulsar \mousepsr. This analysis led to the identification
of distinctive features which should also be observable
in other bow shocks. A comparable system is \pwn\ (also known
as 1SAX~J0617.1+2221), an apparent
X-ray and radio bow shock coincident with the supernova remnant (SNR) IC~443,
as shown in Figure~\ref{fig_wide}
(see also Olbert \etal\ 2001\nocite{ocw+01}, hereafter O2001; Bocchino \& Bykov
2001\nocite{bb01b}).  Pulsations from a pulsar in \pwn\ have not yet been
detected, but the nebula's cometary morphology, with a 
point source
near the apex, make it virtually certain that this is a pulsar-driven
bow shock.  
Here we present a deep \cxo\ observation of \pwn, with which
we study the structure of this source and
compare it to that seen for the Mouse.  

Before discussing this system further, we note that it is unclear if
\pwn\ and \snr\ are genuinely associated because, as can be seen in
Figure~\ref{fig_wide}, the cometary trail of \pwn\ points $\sim80^\circ$
away from the direction expected if the embedded pulsar is moving away
from the SNR's center.  Despite this difficulty, for now we assume that
the pulsar, 
PWN and SNR are physically associated, and consequently adopt a common
distance to them of $1.5d_{1.5}$~kpc \citep{ws03}.  In \S\ref{sec_assoc}
we will discuss the validity of this assumption in the light of the new
data presented here.

\section{Observations and Analysis}

The earlier observations of O2001\nocite{ocw+01} were
carried out using the ACIS-I3 CCD of \cxo\ on 2000 April 10-11
(ObsID 760), with \pwn\ located $3'$
%(\dataset[ADS/Sa.CXO#obs/00760]{ObsID 760}), with \pwn\ $3'$
off-axis.  Our more recent observations were performed using the ACIS-S3
CCD on 2005 January 12-13 (ObsID 5531), 
%CCD on 2005 January 12-13 (\dataset[ADS/Sa.CXO#obs/05531]{ObsID 5531}), 
with the source positioned close to the aimpoint.  We have analyzed
both datasets 
using CIAO v3.3 and CALDB  v3.2.1.  For the data from 2000, 
we applied a
known aspect offset to the data-set, and then reprocessed the event 
list\footnote{This
correction was part of the standard pipeline for the 2005 data.}  
to identify hot pixels and afterglows.
For both data-sets, the latest ACIS gain map, the time-dependent gain
correction, and a correction for charge transfer inefficiency were all
applied. The data were then filtered for bad event grades, good time
intervals and flaring. The final usable exposures were 9.6~ks and
37.5~ks for the 2000 and 2005 data, respectively.

An image from the 2005 data in the energy range 0.3--10~keV
is shown in Figure~\ref{fig_main}(a).  A smoothed version, with
radio contours from O2001 overlaid, is presented in
Figure~\ref{fig_main}(b). The resulting image is $\sim6$ times
deeper than that from the 2000 data of O2001.

For spectroscopy, we combined the two data-sets.  We used
identical source extraction regions for each observation (see
Fig.~\ref{fig_main}[b]), and then created
corresponding weighted response and effective area files using the
CIAO script {\tt specextract}. Spectra were grouped so that there
was a minimum of 25 counts per channel (with the exception of the
2000-epoch spectrum of the central source
CXOU~J061705.3+222127 [see \S\ref{sec_images}],
which was grouped to 15 counts per
channel because of the low number of counts).  For background spectra,
we used annular regions immediately surrounding each of the regions
shown in Figure~\ref{fig_main}(b).  Spectra were then modeled using
XSPEC v11.3.1.  In all spectral fitting, we jointly fit the two
observations from 2000 and 2005, with all model parameters locked
between the two data-sets.

\section{Results}

\subsection{Images}
\label{sec_images}

\begin{figure}
%\centerline{\psfig{file=fig_main_6.eps,width=0.8\textwidth}}
\centerline{\psfig{file=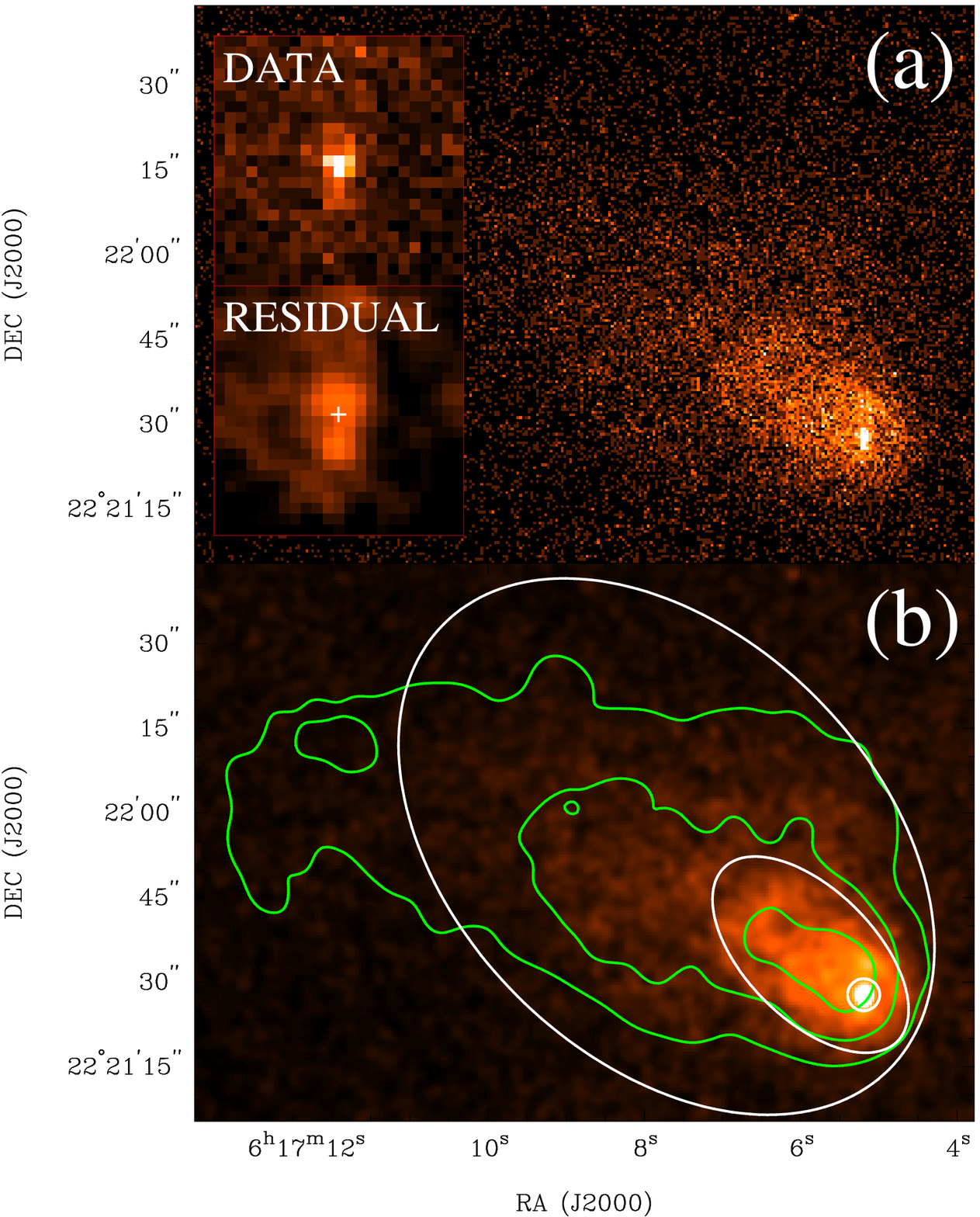,width=0.8\textwidth}}
\caption{X-ray images of the bow shock PWN \pwn\ in SNR~\snr.  (a) A 
\cxo\
image in the energy range 0.3--10~keV, made using the 2005 data and 
displayed using a linear transfer
function ranging from 0 to 9 counts per pixel (the peak value is 85
counts). The insets show the data and the residual after fitting
a central gaussian
to a $11''\times11''$ region surrounding the central source 
CXOU~J061705.3+222127,
both on linear scales ranging from 0 to 40
counts per $0\farcs49\times0\farcs49$ pixel.
The residual image has been
smoothed with a gaussian of FWHM $1''$ to make
clear the existence of faint extended structure on either side of 
CXOU~J061705.3+222127 
(the position of CXOU~J061705.3+222127
is marked with a white cross).
(b) The same data as in panel (a), but smoothed with a gaussian
of FWHM $2''$.
From the inside out, the white ellipses
show the extraction regions for CXOU~J061705.3+222127 (for clarity,
drawn with a radius twice that actually used), the tongue, and the tail;
each region excludes the ones inside it.
The green contours show the radio nebula, using
the same levels as shown 
in Fig.~\ref{fig_wide}.}
\label{fig_main}
\end{figure}

We identify a number of notable features from the X-ray and radio
images shown in Figure~\ref{fig_main}:

\begin{enumerate}

\item As earlier identified by O2001, there is a bright compact source
near the apex of the nebula, which they designated CXOU~J061705.3+222127
(hereafter \psr).  A gaussian fit to this source yields a position
RA~(J2000) $06^{\rm h}17^{\rm m}05\fs18\pm0\fs02$, Decl.~(J2000)
$+22^\circ21'27\farcs6\pm0\farcs3$, where the dominant source of error
is the precision to which we can register the nearby infrared
source 2MASS~06171300+2220587 to its X-ray counterpart.

\item The image in Figure~\ref{fig_main}(a) strongly suggests that
there is extended structure $1''-2''$ to the north and south of
\psr.  We have investigated this by fitting a gaussian with the
dimensions of the point spread function to the emission at the
position of \psr, and then subtracting this from the data.  With
the unresolved component ($\sim200$~counts) removed, the inset to
Figure~\ref{fig_main}(a) demonstrates the clear presence of two
compact components of emission to the north and south of \psr, each
containing $\sim30-40$~counts.

\item On larger scales, an elongated diffuse region with a ``bullet''
morphology is seen, which following from our earlier studies of the
Mouse (G2004), we designate ``the tongue''.  Defining the tongue's
perimeter to be the locus at which the emission falls to 10\% of the
peak in Figure~\ref{fig_main}(b), we find that the tongue is oriented at
a position angle of $+50^\circ$ (N through E), and has an approximate
extent of $33'' \times 15''$. Along its major axis, the tongue extends
$8''\pm1''$ ahead of \psr, and $25''\pm1''$ behind it.  The tongue is
also a feature of enhanced brightness at radio wavelengths (contours
in Fig.~\ref{fig_main}[b]).

\item Surrounding the tongue, we identify a larger diffuse
X-ray region, best seen in Figure~\ref{fig_main}(b), 
which we designate ``the tail''. The tail has a cometary
shape, but with a broader opening angle than the tongue.  It is sharply
bounded at its apex, but fades into the background toward the rear. The
tail is the brightest 
part of a larger X-ray structure seen by \xmm\ \citep{bb01b}.

\end{enumerate}

\subsection{Spectroscopy}
\label{sec_spec}

We have extracted spectra for the three regions shown in
Figure~\ref{fig_main}(b), corresponding to~\psr, the tongue, and
the tail.

To accurately estimate the foreground absorbing column, $N_H$,
we fit simultaneous absorbed power laws to the large
number of counts obtained for the tongue and the tail,
constrained to have the same $N_H$, but with independent photon indices.
The results are given in Table~\ref{tab_spec}, showing that both regions
are good fits to power law spectra, with a foreground column $N_H =
(7.2\pm0.6)\times10^{21}$~cm$^{-2}$ (this and all subsequent errors are
at 90\% confidence). The photon indices for the two regions are the same
within the errors.

\begin{figure}
%\centerline{\psfig{file=fig_spec.eps,width=\textwidth,angle=270}}
\centerline{\psfig{file=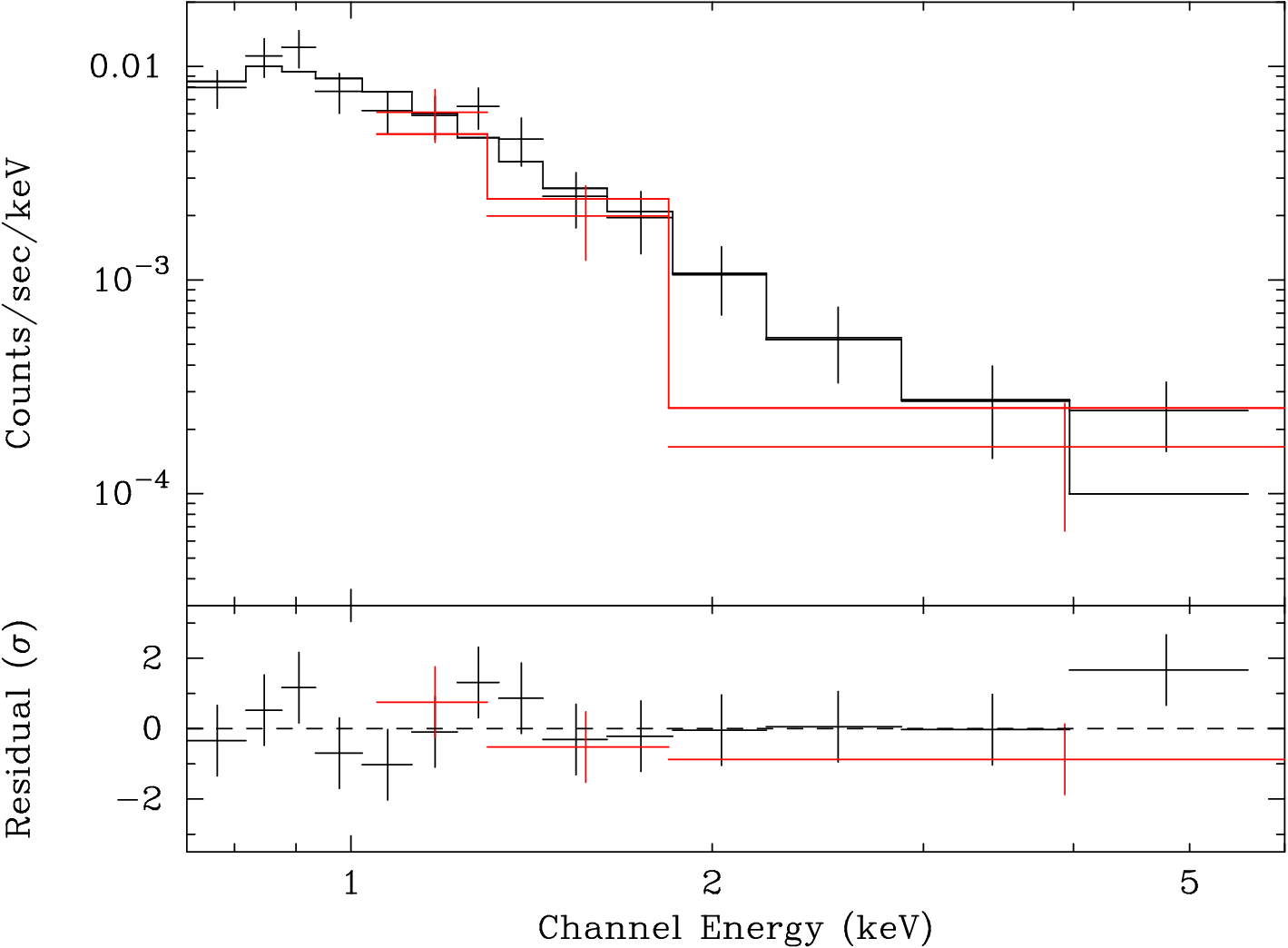,width=\textwidth,angle=270}}
\caption{\cxo\ ACIS spectra of \psr. The points in the upper panel indicate
the data from 2000 (red) and 2005 (black), while the solid lines show
the corresponding best fit absorbed blackbody plus power-law  models. The lower panel
shows the number of standard deviations by which the model and the data
differ in each bin.}
\label{fig_spec}
\end{figure}

We then fit to the spectrum of \psr, excluding 
most of the emission from the compact structures
immediately surrounding it by using a circular extraction region of
radius $1\farcs4$. If we constrain $N_H$ to lie in the range inferred
above, we find that both a power law and a blackbody are poor fits to
the data, both models not accounting for a hard excess seen above 2~keV.
A combined fit of a blackbody and a power law resolves this problem and
produces an excellent fit, as listed in Table~\ref{tab_spec} and shown
in Figure~\ref{fig_spec}.

We have also tried to fit spectra to the compact knots seen on
either side of the pulsar. However, since each knot only contains
$\sim40$ counts in the combined data-set, we are unable to derive
useful spectral parameters for these regions. Simple hardness ratios
indicate that both these knots are significantly harder than \psr,
suggesting that their emission is probably non-thermal.

\section{Discussion}
\label{sec_disc}

Our new data clearly confirm the main result of O2001: \pwn\ consists
of a compact source at the apex
of a striking cometary nebula, which presumably represent an energetic
pulsar and its 
wind-driven bow shock, respectively. Here we focus on the additional information
provided by the higher sensitivity and improved angular resolution of our
new data.

\subsection{The Neutron Star}

We first consider the X-ray emission from \psr.
O2001 assumed that the X-rays from \psr\ were thermal. They
adopted a low absorbing column $N_H = 1.3\times10^{21}$~cm$^{-2}$,
and consequently found a very high temperature ($\approx 0.7$~keV)
and a small emitting area ($\approx0.025d_{1.5}$~km$^2$). With 
better statistics but limited spatial resolution, \cite{bb01b} were
unable to directly detect X-rays from \psr, but found that a blackbody
with a temperature 0.13~keV and radius 3.3~km was consistent with
their data.

Here we directly detect thermal emission from \psr, confirming its
identification as a neutron star. The blackbody component of the fit
to \psr\ shown in Figure~\ref{fig_spec} implies an emitting radius (as
viewed at infinity) of $6^{+8}_{-3} d_{1.5}$~km. This is
not especially constraining, being consistent both with
cooling emission from the whole surface, and also with the 
smaller radii seen for blackbody fits to neutron star atmospheres
\citep[e.g.,][]{pzs+01,mzc+04}.
The inferred temperature and bolometric
luminosity, as viewed at infinity, are $kT^{\infty}
= 102\pm22$~eV
and $L^{\infty} = 5.0^{+5.0}_{-2.2}~d_{1.5}^2 \times10^{32}$~ergs~s$^{-1}$,
respectively.  The age estimated for \snr\ is $\sim30\,000$~years
\citep{che99}.  Assuming that \psr\ is associated with \snr, has a mass
in the range 1.35-1.45~M$_\odot$ and has a 1p proton superfluid core, the
neutron star's predicted  surface temperature is $kT^{\infty} = 60-110$~eV,
and its expected luminosity is $L^{\infty} = 10^{32}-10^{34}$~ergs~s$^{-1}$
\citep[e.g.,][]{kyg02,pgw06}.  Thus, while we are not able to distinguish
between different neutron star 
cooling models, the thermal emission from \psr\ is
consistent with ``standard'' cooling at the age estimated for \snr.

As discussed in \S\ref{sec_spec}, the X-ray spectrum of \psr\ also
requires a significant non-thermal component.
We interpret this as
magnetospheric emission from the pulsar, which typically has a photon
index $\Gamma \sim 2$ \citep[e.g.,][]{pzs+01,mmcb02,jh05}, 
consistent with the fit obtained here.  In this
case, we expect these X-rays to be strongly pulsed, as is seen for several
other young and energetic pulsars \citep[see][for a review]{krh06}.  
This possibility can
be tested through a future observation with the \cxo\ High Resolution
Camera.

\subsection{Compact Structures Close to the Neutron Star}

We will argue in \S\ref{sec_term} that the tongue region in which \psr\
is embedded traces the outer surface of the wind termination shock.
Since the region interior to this corresponds to cold, unshocked wind
material which should not emit,
the compact X-ray structures seen close to the neutron star 
likely lie outside the tongue, but projected against it.

As two emitting components on either side of the pulsar, there
is the temptation to identify these structures as polar jets, as
have been seen for several Crab-like PWNe
\citep[e.g.,][]{wht+00,hgh01}.  Such jets are now understood to be formed
beyond the wind termination shock by equatorially outflowing particles
which reverse in flow direction
and which are then collimated along the spin axis via hoop 
stress \citep{kl03,dab04}.  For the Crab Nebula, the reversal
of the equatorial flow depends on the confining thermal pressure into
which the PWN expands \citep{kl03}.  However, for \pwn, ram pressure
produces a very different set of boundary conditions, for which it is
as yet unclear if such structures can form.
Alternatively, X-ray knots close to the pulsar have been
seen in PWNe such as the Crab Nebula, G320.4--1.2, G292.0+1.8 and the Mouse
\citep[][G2004]{hmb+02,gak+02,hsp+03}.  \cite{kl04} propose that these features
are relativistically beamed components beyond the termination shock,
where the shocked flow travels along a curved surface.

In either interpretation, such structures should define the projected
spin axis of the neutron star, as has been argued for the Crab
pulsar and several other sources \citep{nr04}. If the projected
velocity vector of \psr\ is defined by the bow-shock morphology,
then \pwn\ would represent an example of substantial misalignment
between the pulsar's spin axis and its velocity vector. This would
be in sharp contrast to other systems for which ``spin-kick alignment''
has been claimed, and would have important implications for the
formation mechanism of neutron stars \citep{lcc01}. Alternatively,
as discussed below in \S\ref{sec_assoc}, the pulsar may in fact be
moving to the south, away from the center of SNR~\snr.  In this
case the spin and velocity vectors would show approximate alignment,
and the bow-shock morphology would trace the effect of surrounding bulk
flows rather than of the pulsar's motion.  We are developing full
three-dimensional
hydrodynamic models of bow shocks with anisotropic outflows
to address these possibilities \citep{vcm06}.

\subsection{The Nebula}
\label{sec_neb}

Cometary X-ray nebulae behind moving neutron stars have now
been seen by \cxo\ and \xmm\ for several sources \cite[][and references
therein]{gs06}. However, what our new data make clear is that \pwn\
is clearly structured into two distinct components: the bright
tongue, and the fainter surrounding tail. 
We previously identified this same spatial decomposition for the Mouse,
and adopt a similar interpretation here: namely that the
bullet-shaped tongue region represents the surface of the wind termination
shock, while the tail corresponds to shocked material flowing downstream
(see G2004 for detailed arguments and discussion).

\subsubsection{The Termination Shock}
\label{sec_term}

We directly resolve the separation
between \psr\ and the apex of the tongue, and so can infer a
forward termination shock radius $r_{TS}^F \approx 0.06d_{1.5}$~pc. There
is also a clear termination of the tongue behind \psr,
so we also infer a backward termination shock radius $r_{TS}^B
\approx 0.2\mu d_{1.5}$~pc, where $\mu \le 1$ is a factor which accounts
for smearing of the backward termination shock as a result of ions in
the wind (see \S4.5 of G2004).

The ratio $r_{TS}^B / r_{TS}^F$ is a direct diagnostic of the Mach
number, $\mathcal{M}$, of the system.  For low Mach
numbers, we expect $r_{TS}^B / r_{TS}^F = \gamma^{1/2} \mathcal{M}$, where
$\gamma = 5/3$ is the adiabatic coefficient of ambient gas
\citep{buc02a,vag+03}.  Here
we find  $r_{TS}^B / r_{TS}^F \approx 3\mu$ independent of distance,
from which we infer\footnote{Note that the PWN's opening angle cannot
be used to infer $\mathcal{M}$, as that  only
applies to the unseen forward shock.} $\mathcal{M}
\approx 1.2 $ for $\mu =0.5$ (see G2004). 
This bow shock is only mildly supersonic, in
sharp contrast to the Mouse for which we found $r_{TS}^B / r_{TS}^F
\approx 13\mu$ and $\mathcal{M} \approx 60$. This is consistent
with the expected conditions: the Mouse is moving through the
interstellar medium (ISM) where the sound speed is low, while if \pwn\ is
in the SNR
interior where the sound speed of hot shocked gas is
high, we expect $\mathcal{M} < 3$ \citep[see Fig.~4 of][]{vdk04}.

If \pwn\ and \snr\ are associated, we can combine the observed
properties of these objects to
estimate the space velocity and spin-down luminosity
of \psr.
\cite{kon+02} find that in the outer parts of \snr, the shocked gas
has a temperature 0.2~keV, corresponding to a sound speed $c_s \approx
180$~\kms. Since \psr\ is near the edge of the shell of \snr\ (see Fig.~\ref{fig_wide}), 
we adopt that value here.  The pulsar velocity relative to surrounding gas
is then $V_{rel} = \mathcal{M} c_s\approx 420\mu$~\kms.
To determine the pulsar space velocity, we must add vectorially
the local flow velocity, $V_{flow}$,
to $V_{rel}$.  The PWN
is close to the SNR's rim, so we assume $V_{flow} \sim V_{SNR}
\approx 100$~\kms\  \citep{che99}, where $V_{SNR}$ is the velocity of the
forward shock. 
If the local flow is directed radially outward from
the SNR's geometric center, then $V_{rel}$ and $V_{flow}$
are approximately perpendicular.
If we adopt $\mu=0.5$, we then find a pulsar
velocity $V_{PSR} \approx 230$~\kms. By assuming
that the pulsar was born at the SNR's geometric center
(see discussion of this point in \S\ref{sec_assoc}),
O2001 were able to derive an independent velocity
estimate $V_{PSR}
\approx 250\pm50$~\kms, which is in good agreement with
our calculation.

We can estimate the pulsar's spin-down luminosity, $\dot{E}$,
as follows. If \pwn\ is inside the SNR but close to its edge, the ambient
pressure is $P_0 \approx P_{sh} = 3/4 \rho_0 V_{SNR}^2$, where
$P_{sh}$ is the pressure behind the SNR's forward shock and $\rho_0$
is the density of the ambient ISM.  \cite{che99} gives $\rho_0 \approx
2.5d_{1.5}^{-1} \times 10^{-23}$~g~cm$^{-3}$, so we infer\footnote{A
similar estimate comes from the X-ray spectrum of \snr, which give $P_0 \approx
1d_{1.5}^{-1/2} \times 10^{-9}$~ergs~cm$^{-3}$ \citep{kon+02}.} $P_0
\approx 2d_{1.5}^{-1}\times10^{-9}$~ergs~cm$^{-3}$.  Pressure balance
between the bow shock and its surroundings then implies $\dot{E}/4\pi
{r_{TS}^F}^2 c = \gamma \mathcal{M}^2 P_0$ (G2004).  For $\mu = 0.5$ and
$d_{1.5} = 1$, we find $\dot{E} \approx 5\times10^{37}$~ergs~s$^{-1}$,
which would make \psr\ the most energetic member of the 10--100 kyr old
group of ``Vela-like'' pulsars \cite[see Table 5 of][]{kbm+03}.  We note
that our estimate of $\dot{E}$ is more than an order of magnitude larger
than that found by assuming that the X-ray or radio luminosity of \pwn\
is a fixed fraction of $\dot{E}$ \citep[O2001;][]{bb01b}, illustrating
the large uncertainties in all these indirect estimates of $\dot{E}$.

\subsubsection{The Post-Shock Flow}

Another difference between the Mouse and \pwn\ is the morphology
of the tail.  For the Mouse the tail has a narrow component seen
in both radio and X-rays, enveloped by a broader region seen at
radio wavelengths.  G2004 interpreted this as material
shocked at the backward and forward termination shocks, respectively.
For the latter, the flow speed is $\sim4$ times higher,
but the magnetic field is stronger.  The resultant synchrotron losses explain
its lack of X-ray emission.

In contrast to the Mouse, here the tail is composed of a single
broad structure, with good correspondence between radio and X-rays. This
difference can be understood as follows.  Conservation of energy leads
us to expect that $B_n^F / B_n^B = r^B_{TS}/r^F_{TS}$ (see Eqn.~[13]
of G2004), where $B_n^F$ and $B_n^B$ are the nebular magnetic field
strengths in the flow from the forward and backward termination shocks,
respectively.  In this case we find 
$B_n^F / B_n^B \approx 3\mu$, a ratio
$\sim4$ times smaller than that seen in the Mouse. Factoring in the difference
in flow speed between the forward and backward flow regions \citep{bad05},
the synchrotron loss times for the two zones now become comparable. We
thus expect a single combined region emitting in both the X-ray and
radio bands, as observed.
The power laws of the tongue and tail regions have similar photon indices
(see Table~\ref{tab_spec}), indicating that radiative losses are minimal
over the $\sim0.5$~pc scale imaged by \cxo. At distances $\ga2$~pc
downstream, the X-ray spectrum begins to soften due to synchrotron losses,
as seen by \xmm\ \citep{bb01b}.

\subsection{The Association Between \pwn\ and \snr}
\label{sec_assoc}

As mentioned in \S\ref{sec_intro}, there are difficulties with the
claimed association between \pwn\ and \snr. Most notable is the fact that
the nebular symmetry axis is completely misaligned with respect to
the vector running from the
SNR's center to \psr\ (see Fig.~\ref{fig_wide}). Furthermore, the failure
to detect a period and period derivative for 
\psr\ prevents us from estimating the pulsar's age and comparing
it to 
that of the SNR, while the wide range in $N_H$ seen toward different
parts of \snr\ in previous X-ray studies\footnote{This presumably results from
spatial variations in the absorption due to the well-known interaction of
this SNR with a molecular cloud; see e.g., \cite{bgbw88}.} makes it difficult
to determine whether the PWN and SNR are at similar distances.

However, the results presented here provide new evidence in favor of a
physical association. First, the surface temperature measured for \psr\
is consistent with the age of 30\,000 years inferred for the SNR. Second, as
discussed in \S\ref{sec_neb}, the differences in morphology between \pwn\
and the Mouse imply that either the confining media or the pulsars have
very different properties. Specifically, if we repeat the estimates
made in \S\ref{sec_term}, but now for a pulsar not embedded inside
a SNR but rather moving through a typical warm region of the ISM
($c_s \sim 10$~\kms; $P_0 \sim 3 \times 10^{-13}$~ergs~cm$^{-3}$),
we find $V_{PSR} \approx 20$~\kms\ and $\dot{E} \approx 
10^{34} d_{1.5}^2$~ergs~s$^{-1}$.  The former is slower than 98\% of the
young pulsar population \citep{hllk05}, while the latter would imply a
$\sim10\%$
efficiency of conversion of spin-down luminosity into the
total X-ray luminosity of the system,
which is higher than observed for any other pulsar and its
PWN \cite[cf.,][G2004]{che00}. The properties of the PWN thus argue
that the pulsar wind is most likely confined by hot gas in the SNR interior.

Given these arguments for a genuine association, we then need to
explain the misaligned nebular symmetry axis.  A possible
explanation is that
the explosion site did not coincide with the SNR's geometric
center. This can result if the progenitor star had a high space velocity,
so that the supernova exploded at an offset position
within a wind-blown cavity
\citep{gva02}. The pulsar's velocity vector is then directed away
from the explosion site, rather than from the SNR center.
Note that the offset between the explosion site and the SNR's
center is probably not as extreme as suggested by the observed
morphology, since
gas flows in the SNR interior are likely to be skewing the
orientation of 
\pwn.\footnote{\cite{gjs02} proposed
a similar explanation to account for the misalignment between the
proper motion direction of PSR~J2124--3358 and the symmetry axis of
its surrounding bow shock.}  This can be tested
by future observations: combining the estimates of $V_{rel}$
and $V_{flow}$ made in \S\ref{sec_neb} above, we 
expect a proper motion for \psr\
of $\sim35d_{1.5}^{-1}$~mas~yr$^{-1}$ at
a position angle of $\sim200^\circ$, north through east.
This may be measurable 5--10 years hence. 

%If the pulsar was born
%off-center, it should be moving in the south-westerly direction
%implied by the bow-shock morphology. If the tail is being skewed
%by non-radial flows, the pulsar is likely moving southward, away
%from the center of \snr.

\section{Conclusions}
\label{sec_conc}

Our new \cxo\ observations of \pwn\
reveal a
thermally-emitting neutron star 
with a typical space velocity of
$\sim230$~\kms, whose relativistic wind
drives a bow shock through the shocked gas inside its
associated SNR~\snr. 
We place this system midway between PSR~B1853+01 / SNR~W44 (for which the
pulsar bow shock is deep in the interior of the SNR) and PSR~B1951+32
/ SNR~CTB~80 (for which the pulsar is now crossing the SNR shell) in the
evolutionary sequence described by \cite{vdk04}.
While pulsations from the central source in \pwn\ have not been
detected in the radio band, pulsed magnetospheric emission may be
detectable in future deep X-ray observations.

\pwn\ shows a close correspondence with the Mouse, in that the X-ray
emission from both sources consists of two distinct components, a
``tongue'' and a ``tail''. The specific differences between 
these features in \pwn\ and in the Mouse can be understood in terms of the
much lower Mach number of the pulsar for \pwn. The presence of these
features in two systems, and our ability to interpret them consistently
in terms of the expected wind shock structures, argue that these are
ubiquitous features in pulsar bow shocks.  Deeper observations of other
pulsar bow shocks \cite[especially those with detected
pulsars and known proper motions, see][]{gjs02,sgk+03,cbd+03}, 
in hand with new relativistic magnetohydrodynamic
simulations of these systems \cite[e.g.,][]{bad05}, can further add to
our understanding of these systems.

\begin{acknowledgements}

We thank Denis Leahy for providing a radio image of \snr, and an anonymous
referee for comments which improved this paper.  The Second Palomar
Observatory Sky Survey (POSS-II) was made by the California Institute of
Technology with funds from the National Science Foundation, the National
Aeronautics and Space Administration, the National Geographic Society,
the Sloan Foundation, the Samuel Oschin Foundation, and the Eastman Kodak
Corporation.  The Oschin Schmidt Telescope is operated by the California
Institute of Technology and Palomar Observatory.  This paper has made use
of NASA's {\em SkyView}\ facility ({\tt http://skyview.gsfc.nasa.gov})
located at NASA Goddard Space Flight Center.  
NRAO is a facility of the National Science Foundation operated under
cooperative agreement by Associated Universities, Inc.
B.M.G. acknowledges the
support of NASA through Chandra grant GO5-6052X and LTSA grant NAG5-13032.
P.O.S. acknowledges support from NASA contract NAS8-39073.
F.C. acknowledges the support of the National Science Foundation
through grant AST-05-07376.

\end{acknowledgements}

{\it Facilities:} CXO (ACIS), VLA

%\bibliographystyle{apj}
%\bibliography{journals,modrefs,psrrefs,crossrefs}

%\clearpage

\begin{deluxetable}{ccccccc}[bht]
\tablecaption{Spectral fits to regions of \pwn.\label{tab_spec}}
\tabletypesize{\scriptsize}
\tablehead{Region & Total Counts\tablenotemark{a} &
$N_H$                 & $\Gamma$ & $kT^\infty$ & $f_x$  & $\chi^2/\nu$ \\
       &             (0.5--10.0~keV)        & ($10^{21}$~cm$^{-2}$) &
& (eV) & ($10^{-13}$~ergs~cm$^{-2}$~s$^{-1}$)\tablenotemark{b}}
\startdata
  Tongue (PL)   &   $4432\pm96$ &    $7.2\pm0.6$\tablenotemark{c} 
   & $1.70^{+0.10}_{-0.05}$ &
--- & $17\pm1$ & 447/511 = 0.88 \\
  Tail (PL)  &   $6999\pm119$ &        ''             & $1.73\pm0.08$ & 
--- & $27\pm2$ &  '' \\
  Central source (PL)  &   $313\pm20$ & $6.6-7.8$ (constrained) &
$5.6^{+0.7}_{-0.6}$ & --- &  $0.47^{+0.32}_{-0.09}$ & 25/14 = 1.79 \\
  Central source (BB)  &   '' & ''  & --- &
$150\pm20$        & $0.21^{+0.15}_{-0.10}$ & $46/14 = 3.31$ \\
  Central source (BB+PL)  &   '' & '' & $2.6^{+0.5}_{-1.0}$ &
$102\pm22$        & $0.45^{+0.46}_{-0.22}$ & $10/12 = 0.85$ \\
\enddata
\normalsize

\tablecomments{Uncertainties are all at 90\% confidence. ``PL'' = power law
and ``BB'' = blackbody.}

\tablenotetext{a}{The number of counts is that for
the 2000 and 2005 observations combined. In each case,
the 2005 data contribute $\sim$85\%--90\% of the total.}

\tablenotetext{b}{Fluxes are for the energy range 0.5--10.0~keV,
and have been corrected for absorption.}

\tablenotetext{c}{The spectral fits for the tongue and tail regions were
constrained to have the same value of $N_H$.}

\normalsize
\end{deluxetable}

%\clearpage

\clearpage

\clearpage

\end{document}